\documentclass[manuscript,screen,review=false]{acmart}

\usepackage{subcaption}
\usepackage{algorithm}
\usepackage[noend]{algpseudocode}

\usepackage{ulem}

\usepackage{graphicx}
\usepackage{subcaption}
\usepackage{booktabs} 
\usepackage{pdflscape} 
\usepackage{afterpage}
\usepackage[longtable]{multirow}
\usepackage{longtable}
\usepackage[nointegrals]{wasysym}
\usepackage{arydshln}

\AtBeginDocument{%
  \providecommand\BibTeX{{%
    \normalfont B\kern-0.5em{\scshape i\kern-0.25em b}\kern-0.8em\TeX}}}

\setcopyright{acmcopyright}
\copyrightyear{2023}
\acmYear{2023}
\acmDOI{XXXXXXX.XXXXXXX}

\begin{document}

\title[Workflow Scheduling in Cloud and Edge Computing Environments with Reinforcement Learning]{Reinforcement Learning based Workflow Scheduling in Cloud and Edge Computing Environments: A Taxonomy, Review and Future Directions}
\author{Amanda Jayanetti}
\affiliation{%
 \institution{The University of Melbourne}
 \country{Australia}}

\author{Saman Halgamuge}
\affiliation{%
 \institution{The University of Melbourne}
 \country{Australia}}

 \author{Rajkumar Buyya}
\affiliation{%
 \institution{The University of Melbourne}
 \country{Australia}}

\renewcommand{\shortauthors}{Jayanetti, Halgamuge, Buyya }

\begin{abstract}
Deep Reinforcement Learning (DRL) techniques have been successfully applied for solving complex decision-making and control tasks in multiple fields including robotics, autonomous driving, healthcare and natural language processing. The ability of DRL agents to learn from experience and utilize real-time data for making decisions makes it an ideal candidate for dealing with the complexities associated with the problem of workflow scheduling in highly dynamic cloud and edge computing environments. Despite the benefits of DRL, there are multiple challenges associated with the application of DRL techniques including multi-objectivity, curse of dimensionality, partial observability and multi-agent coordination. In this paper, we comprehensively analyze the challenges and opportunities associated with the design and implementation of DRL oriented solutions for workflow scheduling in cloud and edge computing environments. Based on the identified characteristics, we propose a taxonomy of workflow scheduling with DRL. We map reviewed works with respect to the taxonomy to identify their strengths and weaknesses. Based on taxonomy driven analysis, we propose novel future research directions for the field. 
\end{abstract}




\maketitle

\section{Introduction}

Workflow is an application model that facilitates the representation of data in a distributed and structured manner. Therefore, the workflow application model is used for modeling the computations and data dependencies of a wide variety of applications ranging from scientific applications that are used in astronomical, biological and medicinal fields to commercial applications that are used in emerging fields such as IoT (Internet of Things). 

Cloud computing has emerged as an efficient computing platform used for provisioning services over the internet. Cloud computing offers a multitude of benefits to users, including on-demand access to a pool of shared resources, rapid elasticity to suit demand variations, reduced start-up costs, elimination of infrastructure management overhead, and so on. Owing to the aforementioned benefits, cloud computing environments are widely used for the execution of workflows. Workflow scheduling in cloud computing is a well studied topic. A plethora of heuristic and meta-heuristic approaches have been proposed in academia for enhancing the performance of scientific workflow executions in cloud computing environments with respect to different optimization objectives including makespan, cost, energy-efficiency and so on. 

Recently, distributed computing paradigms such as multi-clouds and edge computing have emerged to complement the traditional cloud computing paradigm.  Multi-cloud platforms enable the users to consume services from multiple vendors. Whereas, Edge computing complements the cloud by extending computing resources to the network edge thus allowing data to be processed closer to where it is generated. While these new computing paradigms significantly enhance the capabilities of traditional cloud computing offerings with reduced latency, improved security, increased cost-effectiveness and so on, a number of new challenges are also introduced. In order to leverage the full potential of these technologies, adaptive and computationally efficient workflow scheduling algorithms that are capable of satisfying diverse optimization goals amid highly dynamic conditions are needed. 

Reinforcement Learning (RL) is a branch of machine learning that is based on autonomous and experience-driven learning. The RL agent learns through feedback received from its interactions with the environment, and thereby improves its capabilities for making desirable decisions over time. This leads to an intelligent autonomous agent that can adapt to changing conditions.  Despite the benefits of RL, the application of RL techniques to high-dimensional real-world problems has been constrained by limitations such as lack of scalability and curse of dimensionality. The emergence of Deep Learning led to a significant advancement in multiple areas of machine learning resulting in a dramatic improvement in tasks such as voice recognition, language processing and object detection. The combination of deep learning and reinforcement learning gave rise to the field of Deep Reinforcement Learning (DRL).  DRL significantly accelerated the progress of the traditional RL paradigm through the use of deep neural networks for function approximation. The reduction of memory and computational complexities enabled the application of DRL to problems that were formerly intractable with RL. This survey focuses on reviewing RL and DRL techniques that have been proposed for solving the  challenges associated with scheduling workflows in cloud and edge computing environments.

The rest of the paper is organised as follows: In section \ref{survey-background}, a background related to the areas covered in the survey is provided. Section \ref{survey-taxonomy} presents the proposed taxonomy and Section \ref{survey-review} reviews the works surveyed based on the proposed taxonomy. Finally, in Section \ref{survey-future}, potential future areas that can be explored in the area of workflow scheduling with DRL is presented.

\section{Background}\label{survey-background} 

\begin{figure*}[!t]
	\centering
	\includegraphics[width=0.7\textwidth]{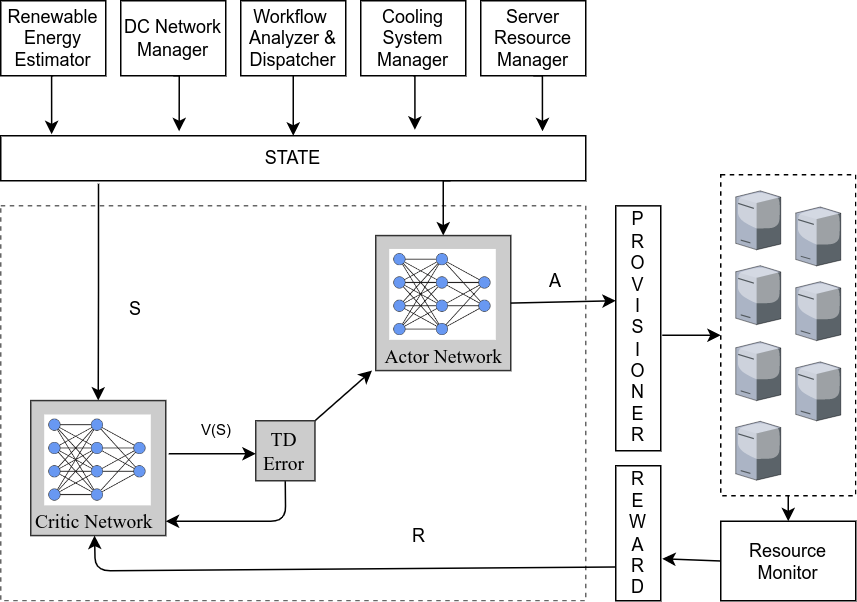}
	\caption{A potential architecture for a holistic workflow scheduling framework in a cloud datacenter with DRL (Actor-Critic method)}
	\label{fig:survey-drl}
\end{figure*}

\subsection{Cloud and edge computing environments}

Cloud services are provisioned to users through three delivery models: Software as a Service (SaaS), Platform as a Service (PaaS) and Infrastructure as a Service (IaaS), and users are charged for the utilisation of these services following a pay-as-you-go model. In SaaS delivery model, applications are offered to users over the internet in a fully managed manner. This eliminates the need for users to engage in downloading, installing, managing or maintaining the applications by themselves. PaaS delivery model offers a computing platform for the development of software. It relieves application developers from the burden of having to manage underlying computing infrastructures, thus allowing them to fully focus on software development activities. In IaaS delivery model, users are offered access to a pool of computing resources (servers, networking, storage etc.) that can be scaled up and down based on demand. With IaaS, users need not maintain or manage the infrastructures physically, but they are responsible for managing certain aspects such as middleware, operating systems, storage etc. 

As opposed to the centralized cloud, edge computing is a fully decentralized computing architecture that brings computation closer to the sources of data generation. In essence, edge computing aims at leveraging the computing, networking, and storage resources of any device (e.g. routers, switches, access points, base stations, nano data centers, etc.) that resides between the cloud and terminal devices to provision cloud-like services with minimal latency. In addition to low latency, the distributed computing architecture of edge computing allows it to facilitate real-time interactions, mobility support, location awareness, and widespread geographical coverage.

\subsection{Workflow scheduling}

The problem of scheduling workflows in a distributed system is NP-Complete in the general case The objective of the traditional workflow scheduling problem is to find an allocation of tasks to execution nodes such that the precedence constraints are satisfied and the workflow completes execution within the minimum completion time. The problem of scheduling workflows in a distributed system is NP-Complete in the general case \cite{ullman1975np}. Scheduling workflows across highly dynamic multi-cloud and edge computing environments adds a further layer of complexity atop the general problem of workflow scheduling across distributed computing environments. From a consumer point of view, more efficient workflow scheduling strategies are needed for satisfying the delay requirements of executions while also minimizing the monetary cost associated with leasing the underlying infrastructures from vendors. Whereas from a vendor perspective, efficient scheduling strategies are needed for distributing consumer workflows across their infrastructures such that the cost of operations including energy consumption of the infrastructure is minimized while also satisfying the Quality of Service (QoS) guarantees. 

Workflow schedulers operating in distributed computing environments primarily focus on mapping tasks to nodes for execution while ensuring data dependencies are satisfied. Depending on the context of the problem the scheduling algorithm may attempt to optimize one or more performance objectives including energy, makespan, cost and QoS (Quality of Service). Scheduling algorithms can be categorised as static (offline) or dynamic (online) depending on their mode of operation. Static algorithms analyze the whole workflow and resource capacities and create a fixed schedule (i.e. task to node mapping) prior to the commencement of workflow execution. Dynamic algorithms operate based on information available at runtime. In the following sub-sections, we have briefly reviewed some of the relevant works that have proposed heuristic and meta-heuristics-based algorithms for scheduling workflows across cloud and edge computing environments. 

\subsubsection{Heuristic methods}

A wide variety of heuristics are proposed in academia for scheduling workflows across cloud computing environments with different optimization goals. HEFT \cite{topcuoglu2002performance} is one popular heuristic designed for makespan minimization in static workflow scheduling. This heuristic is used as a basis for task ordering in many workflow scheduling and resource provisioning algorithms. For instance, W. Zheng et. al \cite{zheng2015adaptive} used HEFT for deriving an initial task to host mapping, post to which DVFS (Dynamic Voltage and Frequency Scaling) technique \cite{dayarathna2015data} was applied to selected hosts in a greedy manner. DVFS technique has been used in multiple other studies for energy-efficient workflow scheduling in cloud computing environments \cite{stavrinides2019energy}, \cite{taghinezhad2022energy}. In \cite{xu2015enreal} a migration-based resource allocation policy for energy-efficient workflow scheduling across cloud VMs is proposed. Researchers have also proposed heuristics based on VM scaling \cite{wen2019energy} and task merging \cite{li2015cost} for scheduling workflows with the objectives of enhancing energy efficiency and cost, while meeting deadline constraints. 

A number of heuristics are proposed for scheduling workflows in cloud and edge environments \cite{ijaz2021energy}, \cite{kaur2022multi}.
In \cite{ijaz2021energy} S. Ljaz et. al proposed a heuristic algorithm for energy and makespan-optimized scheduling of workflows across cloud-edge environments. Firstly, tasks ranked based on optimistic processing times are enqueued to a priority queue. For selecting nodes, a weighted bi-objective function based on minimum completion time and energy consumption is formulated, and tasks are allocated to the node with the minimum value. Deadline aware stepwise frequency scaling approach is also incorporated for further minimizing energy consumption by leveraging the unused time slots between multiple tasks scheduled on the same node.

\subsubsection{Meta-heuristic methods}

Meta-heuristics are widely used in the design of scheduling algorithms in cloud computing environments. Particle Swarm Optimization (PSO) is a popularly used meta-heuristic in cloud workflow scheduling. For instance, in \cite{rodriguez2014deadline} PSO is used for scheduling workflows with the objective of minimizing execution costs without violating deadline constraints. Z. Li \cite{li2016security} also used PSO for designing a workflow scheduling algorithm for minimizing workflow execution cost while adhering to deadline and security constraints. In \cite{ismayilov2020neural}, an artificial neural network was integrated with NSGA-II meta-heuristic algorithm to develop a scheduling algorithm that is capable of adapting to the inherent dynamicity associated with cloud computing environments.

Meta-heuristics techniques are increasingly used for designing workflow scheduling algorithms across cloud and edge/fog computing environments. 
P. Hosseinioun et. al \cite{hosseinioun2020new} used an invasive weed optimization and culture evolutionary algorithm for ordering tasks with precedence constraints prior to allocating them for execution in DVFS-enabled fog nodes. The authors have leveraged slack times of tasks for setting low frequencies to processing nodes without compromising pre-defined deadlines to achieve considerable energy savings.  M. Mokini et. al \cite{mokni2023multi} also ordered tasks taking into account the precedence relations with the use of a Topological Sort Algorithm  \cite{ajwani2012topological}. Afterward, Genetic Algorithm was used for scheduling tasks with the objectives of minimizing makespan and cost while optimizing resource utilization. In a work by  C. Wu et. al \cite{wu2018hybrid} workflows are partitioned to two parts, only one of which is offloaded to the fog and cloud tiers. The other partition remains in the terminal layer in which IoT devices are residing. Partitioning is done for the purpose of minimizing data transfer among terminal and fog/cloud layers which is important for reducing energy-consumption as well as network traffic. The proposed approach uses EDA (Estimation of Distribution Algorithm) technique for scheduling the tasks among fog and cloud tiers with the objectives of minimizing energy consumption and makespan. Y. Xie et. al \cite{xie2019novel} used PSO for scheduling workflows in a hybrid cloud and edge computing environment with the objectives of minimizing execution cost and makespan. A weighted bi-objective function is used to establish the desired balance between makespan and cost. In \cite{bacanin2022modified}, N. Bacanin et. al improved the limitations of firefly meta-heuristic algorithm by the incorporation of genetic operators and a learning prodedure based on quasi-reflection. The enhanced algorithm was then used for scheduling workflows across cloud and edge computing environments with the objectives of minimizing makespan and cost.

\subsection{Reinforcement learning}


Despite the satisfactory results achieved by heuristic-based workflow scheduling algorithms, there are inherent limitations associated with heuristic techniques. The design of a heuristic requires advanced domain expertise and may require the collaboration of multiple experts. Regardless of the level of domain expertise possessed by the experts, the fact that heuristics are designed by humans makes them prone to human errors. Furthermore, the chosen heuristics are not guaranteed to be optimal and the selected heuristics could produce results that are far from the optimal achievable results. Although scheduling algorithms based on meta-heuristics are typically capable of overcoming the aforementioned limitations of heuristics, they are less appropriate for highly dynamic cloud environments due to high computational costs and time complexities. Several studies have integrated meta-heuristics with popular heuristics such as HEFT for developing efficient workflow scheduling algorithms \cite{choudhary2018gsa}, \cite{ahmed2021using}.

Owing to the ability of RL-oriented algorithms for identifying optimal behaviors in highly dynamic and unpredictable environments, researchers are increasingly using RL for solving a wide variety of resource management problems in cloud and edge computing environments. Figure \ref{fig:survey-drl} shows a high-level vision for a holistic DRL based scheduling framework that can be adopted in a cloud dataceneter. For the formulation of state of the environment, data from multiple sources can be collected. Real time renewable energy availability as well as predicted energy levels can be acquired from the renewable energy estimator. The network manager provides the current status of the datacenter network whereas cooling system manager provides details on thermal aspects. The utilization levels of the servers are provided by the resource manager. The state of pending workflows as well as workflows under execution is provided by the workflow analyzer. Dispatching component is responsible for managing the execution order of workflows, and for each ready task dispatched, the DRL scheduler selects the node for allocation. The selected action is executed by provisioner, and the outcome of the action is evaluated and provided as reward to the DRL model.

Resource scaling is a fundamental characteristic of the cloud computing paradigm that differentiates it from previous utility computing infrastructures such as grids. Dynamic application (and resource) scaling requires careful consideration of multiple factors including network traffic and QoS (Quality of Service) requirements of underlying applications. The power of RL is leveraged in multiple works for designing efficient auto-scaling algorithms for cloud applications executing in traditional VMs \cite{barrett2013applying} and spot instances \cite{lin2022learning}, as well as more recent serverless computing environments \cite{schuler2021ai}. VM scheduling and migration is another area that seems to benefit from the capabilities of RL.  
N. Liu et. al \cite{liu2017hierarchical} proposed a hierarchical RL framework comprising a global tier of VM allocation and a local tier of Dynamic Power Management (DPM) \cite{benini2000survey} of servers for optimizing the energy-efficiency of the underlying cloud computing environment. RL was also used by B. Wang \cite{wang2021energy} for designing a VM scheduling and migration algorithm for cloud datacenters. 

RL is increasingly becoming a popular candidate for resolving the resource management problems of upcoming edge and fog computing environments. Q learning was used in multiple works for designing load-balancing algorithms for fog computing environments \cite{talaat2020load}, \cite{baek2019managing}. In \cite{gao2019deep}, Deep Q Learning is used for handling mobility-aware VM migration in edge computing. Deep Q learning was also used in \cite{liu2019deep} for IoT network clustering. 


\begin{figure}[!t]
	\centering
	\includegraphics[width=\textwidth]{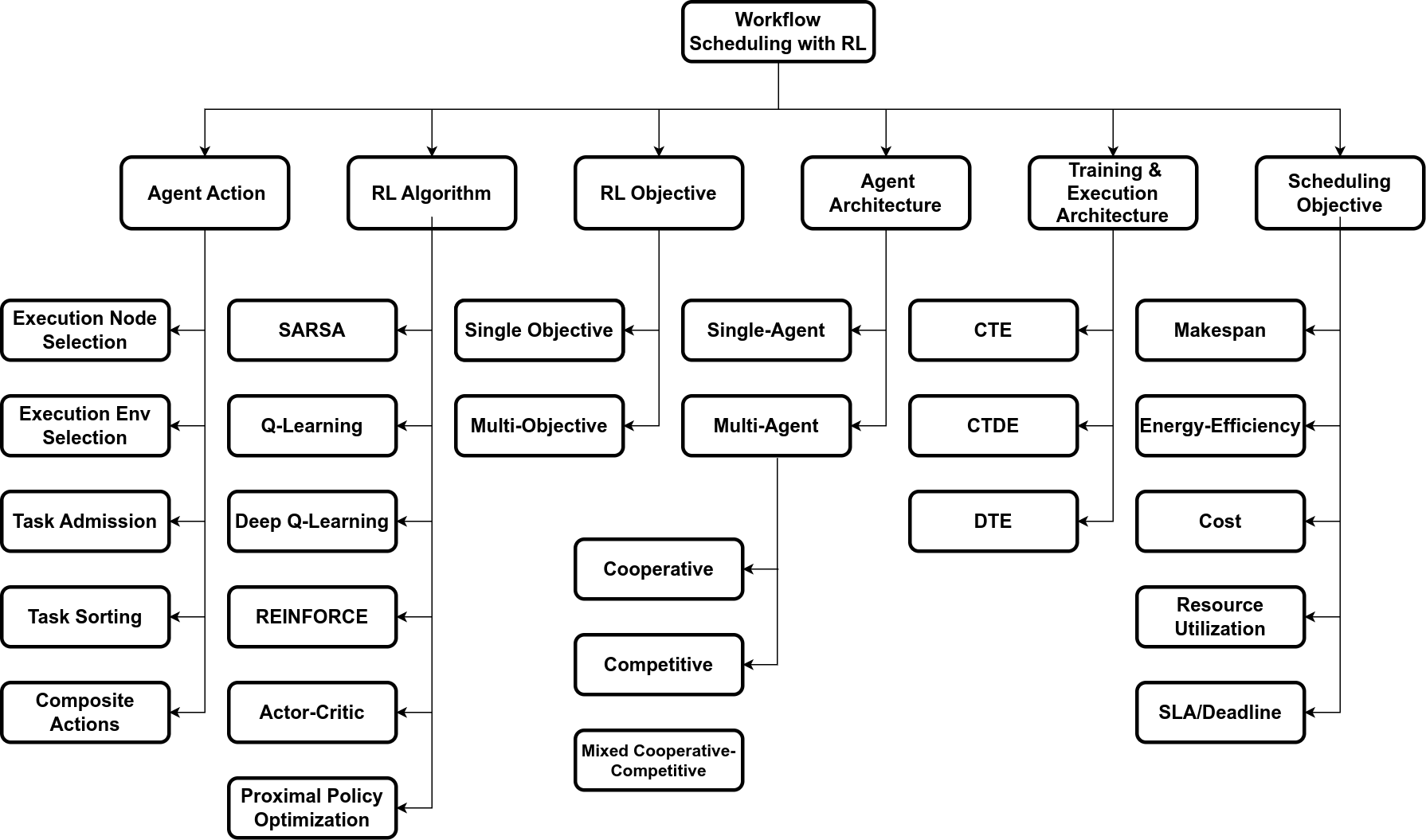}
	\caption{Taxonomy of Workflow Scheduling in Cloud and Edge Computing Environments with Reinforcement Learning [CTE - Centralized Training and Execution, CTDE - Centralized Training and Distributed Execution, DTE - Distributed Training and Execution]}
	\label{fig:taxonomy}
\end{figure}

\section{Taxonomy}\label{survey-taxonomy}

\subsection{Taxonomy based on the specific problem for which RL is used}

In this section, we analyse the Agent Action branch of the taxonomy shown in Figure \ref{fig:taxonomy}. Reinforcement learning algorithms are used in workflow schedulers to perform many different types of actions. An action as previously discussed results in the state of the environment being transitioned from the current state to the next state, and the agent receives either an immediate reward or a delayed reward (usually at the end of the episode) which reflects the desirability of the action with respect to the objective of the scheduling algorithm.

An important pre-requisite for the successful application of RL in problem-solving is to identify which areas of the problem being solved would benefit from the advanced capabilities of the RL paradigm. In the case of workflow scheduling across distributed computing infrastructures, the most straightforward and common use of RL is the selection of a node for task execution, however, authors in surveyed literature have explored the power of RL in several other areas of workflow scheduling as discussed in the sub-sections below.

\subsubsection{Execution Node Selection} 
In a majority of existing studies, RL is used for determining the node to which a task can be allocated for execution. Such a decision requires the consideration of multiple factors including system objective, specific resource requirements of the task which is to be allocated to the selected execution node, the current resource utilization levels of the node as well as the size of the waiting queue. A well-trained RL agent is capable of finding the right match by taking into account all the aforementioned factors and therefore, RL is heavily used for selecting the node for task executions in workflow schedulers.  

The size of the action space is a parameter that influences the training speed of the algorithm as well as the decision speed \cite{dulac2015deep}. Therefore, in some works, multiple actions are permitted in a single time step so that the size of the action space is constrained. 

\subsubsection{Execution Environment Selection} 
The selection of the execution environment is a high-level action for which RL is used in hierarchically designed schedulers and it is often combined with a lower-level action such as execution node selection. In schedulers that are operating across multi-cloud environments, an action of an RL agent corresponds to the selection of a particular datacenter for allocation of a task, or an entire workflow. In a cloud datacenter where execution nodes are grouped into clusters, an action would be the selection of a server cluster for task allocation. For schedulers that operate across cloud and edge computing environments, an action corresponds to the selection of the particular layer (cloud/edge) to which the task will be allocated \cite{jayanetti2022deep}.  

\subsubsection{Task Admission} 

Task admission includes the selection of a task to be scheduled. In a multi-tenant workflow scheduling system at a particular time step, there may be multiple tasks that have their precedence constraints satisfied, and therefore ready to be scheduled. In such scenarios, the selection of a task from multiple ready tasks is an important factor that impacts the resource utilization of the underlying system as well as the makespan of the workflows that are submitted to the system. Accordingly, in order to perform the admission of a task while taking into account the system objective, resource utilization levels as well as user-defined constraints such as deadlines, the power of RL is leveraged in multiple studies. In \cite{hu2019learning}, an action of the RL agent corresponds to the selection of a task to be scheduled in the cluster. The agent is also allowed to select a movement action which causes the system to run one more time-step (without admitting a new task to the system). Similar to the case in which RL is used for the selection of an execution node, in task admission as well it is important to limit the size of action space to prevent the impediment to the training process. For instance, if there are N tasks to be scheduled at a certain point, the size of the action space would be $2^N$. By allowing the agent to take multiple actions during a single time step the size of action space can be kept linear in N. This approach is adopted in \cite{hu2019learning} to prevent the expansion of action space, thus facilitating the training process. In \cite{asghari2021task}, RL is used for task selection (i.e. admission) in a scenario where an RL agent is assigned to each resource, and the action of the agent is to select a task to be executed on the relevant resource. 

\subsubsection{Task Sorting}

As opposed to independent tasks, a workflow consists of multiple tasks with complex data inter-dependencies. Depending on the structure of the workflow, completion of certain tasks may be more crucial for minimizing the makespan of the workflow. Ordering tasks of a workflow that is to be executed in a heterogeneous platform is an NP-complete problem and therefore a number of heuristics have been proposed to find approximate solutions for this problem \cite{topcuoglu1999task}. HEFT is one such heuristic that is popularly used for task prioritization in cloud workflow schedulers \cite{topcuoglu1999task}. HEFT is a greedy scheduling algorithm which evaluates the lengths of critical paths of tasks and ranks them accordingly, and then, the best processor is allocated to the highest ranked task yet to be scheduled. It is a static method since the workflow is processed and scheduling decisions are made offline. In QL-HEFT authors combine RL with HEFT algorithm for designing a workflow scheduling algorithm capable of minimizing makespan \cite{tong2020ql}. In the proposed approach, the upward ranks of tasks computed according to the HEFT technique are used as immediate rewards for training the RL agent. A. Kaur \cite{kaur2022deep} used Deep Q learning coupled with HEFT heuristic for ordering the tasks of workflows which seemed to have produced significantly better results compared to QL-HEFT. Despite improved makespan guarantees, static scheduling algorithms are not suitable for scheduling workflows across the highly dynamic cloud and edge computing environments. \cite{asghari2020online, asghari2020cloud} proposed online scheduling algorithms for ordering tasks of workflows using RL. In these approaches, an RL agent assigned to the workflow learns to assign values to nodes through multiple rounds of training and the values are then used for task prioritization prior to resource provisioning phase. In \cite{asghari2020cloud}, multiple agents are assigned to parts of the workflow for speeding up the learning process involved in task prioritization. This is particularly useful when it comes to large workflows with complex inter-dependencies.

\subsubsection{Composite Actions}

In a majority of surveyed works, one of the aforementioned single discrete actions is the output of the RL agent. However, such action space designs are prone to the 'curse of dimensionality as the system becomes more complex and the output of the RL agent becomes multi-dimensional leading to a combinatorial increase in the number of potential actions. More advanced action space designs involving composite actions are then useful for preventing the explosion of action space leading to in-efficient learning \cite{tavakoli2018action}. Hierarchical RL is another means of designing RL models such that a more complex high-level problem is decomposed into multiple sub-problems thus restricting the number of discrete actions that need to be considered by lower-level agents \cite{barto2003recent}. One such work that leveraged the power of hierarchical RL is \cite{jayanetti2022deep} in which the authors presented the design of a hierarchical action space for task allocation in edge-cloud computing environments. In the proposed hierarchical design, the upper-level agent is responsible for determining if a task should be executed in cloud or edge, and the lower-level agent then decides which node in the selected tier should be selected for task execution. 

\subsection{Taxonomy based on RL algorithm}

In this section, we discuss the RL algorithm branch of the taxonomy shown in Figure \ref{fig:taxonomy}. Reinforcement learning algorithms can be broadly classified into two categories; Model-based reinforcement learning algorithms and model-free reinforcement learning algorithms. As the name implies model-based algorithms rely on a predictive model of the environment which provides the state transition probabilities and rewards, thus enabling agents to make informed decisions and plan ahead. However, it is impractical to formulate predictive models as such for highly dynamic cloud computing environments. Accordingly, all the surveyed RL algorithms used in cloud workflow scheduling fall into the model-free category in which an agent learns completely through the experience it gathers by interacting with the environment. A number of popular RL algorithms are considered in this section of the taxonomy.  

\subsubsection{SARSA}

SARSA is a Temporal Difference based RL algorithm that uses the following equation for learning the Q values after every state transition:

\begin{equation}
Q(s,a) = Q(s,a) + \alpha [R(s,a,s') + \gamma Q(s',a') - Q(s,a)] 
\end{equation}

Since the behavior policy is updated based on the actions taken and rewards received, SARSA is an on-policy RL algorithm.

\subsubsection{Q Learning}

Q learning is a popular TD (Temporal Difference) control based off-policy RL algorithm which starts with arbitrary Q values and iteratively uses the update rule in equation 1 for converging to the optimal Q value function ($Q^*(s,a)$). 

\begin{equation}
Q(s,a) = Q(s,a) + \alpha [R(s,a,s') + \gamma \max_{a'\in A}Q(s',a') - Q(s,a)] 
\end{equation}

The optimal Q function can then be used for deriving the optimal policy as denoted in the equation below.

\begin{equation}
\pi^* (a|s) = \max_{a\in A} Q^{*}(s,a)
\end{equation}

However, Q learning relies on storing data in a tabular format which limits its suitability for complex environments with high dimensional state spaces due to space constraints associated with storing experiences as well as the in-feasibility for an agent to explore all states.

\subsubsection{Deep Q Learning} 
DQN is a well known off-policy reinforcement learning algorithm which leverages a neural network for approximating the Q-value function. Neural network is initialized with arbitrary weights which are updated during the course of training thus allowing the network to learn the actual Q value function. This is done by iteratively minimizing the mean squared difference between network predicted Q values and target Q values as shown in equation \ref{dqn}. The parameterized Q-value function can be represented as: $Q(s,a | \theta)$.

\begin{equation}
\label{dqn}
\begin{aligned}
L(\theta) = \frac{1}{N} \sum_{i \in N} (Q(s_i,a_i|\theta) - (Q'(s_i,a_i|\theta))^2 \\ 
where \; (Q'(s_i,a_i|\theta) = R(s_i,a_i) + \gamma \max_{a'\in A}Q(s_i',a_i'|\theta)
\end{aligned}
\end{equation}

The neural network, takes as input the state of the environment ($s\in S$) and outputs the Q value of each action ($a \in A$). The action with maximum Q value can then be selected to derive the optimal policy as follows:

\begin{equation}
\pi^* (a|s,\theta) = \max_{a\in A} Q^{*}(s,a|\theta)
\end{equation}

\subsubsection{REINFORCE} \cite{williams1992simple} is a fundamental policy gradient algorithm that uses a trajectory sampled using the current policy for obtaining the cumulative discounted reward, $v_t$ which in turn is used as an unbiased estimate of $Q(s_t,a_t)$. The policy parameters are updated with gradient ascent. Since the update takes place after a complete trajectory, REINFORCE is an off-policy algorithm. Owing to the differences between sampled trajectories, the gradient estimates could have a high variance giving rise to slow convergence. A baseline function deducted from the return is used as a means of reducing variance in gradient estimates. 
 
\subsubsection{Actor-Critic}

Actor critic methods are a branch of policy gradient algorithms which combine the benefits of value based and policy based RL through the use of two interacting functions termed actor and critic. The actor takes the state of the environment as input and outputs the probability distribution of the actions to be followed for achieving a particular objective. The Critic provides an evaluation of the desirability of the action taken with respect to the set objective, and the feedback is used by actor for updating the policy parameters in the direction suggested by the critic. In essence, the actor learns the policy and critic learns the value function. Neural networks are typically used for parameterizing both functions. 

\subsubsection{Proximal Policy Optimization} \cite{schulman2017proximal}
Vanilla policy gradients suffer from sample inefficiency since a sample is only used once for updating the policy. The PPO method enables performing multiple epochs of updates with mini-batches of samples through the use of a clipped surrogate objective function. Since large policy updates could lead to sub-optimal convergence, the objective function of PPO restricts the degree to which new policy is allowed to deviate from the old policy by clipping the ratio between current policy and new policy in a range, [1 - $\epsilon$, 1 + $\epsilon$].

\subsection{Taxonomy based on RL objective}

In this section, we analyse the RL objective branch of the taxonomy shown in Figure \ref{fig:taxonomy}.
Workflow scheduling problems in cloud and edge computing environments for which RL is used could have a single objective or multiple objectives that may or may not be of equal importance. 

\subsubsection{Single Objective RL (SORL)}

A majority of RL-based workflow schedulers that have been proposed in academia belong to the single objective category. 
Most of the existing single-objective RL algorithms have attempted to optimize scheduling objectives such as energy, makespan, cost, SLA, and resource utilization. An in-depth discussion of these objectives is included in section \ref{section-objective}. It is realistically impossible to map the diverse and conflicting requirements of workflow schedulers deployed in highly dynamic cloud and edge computing environments to a single objective scheduling problem. Therefore, a popular technique is to use scalarization for transforming the multi-objective scheduling problem into a single objective RL problem with the use of a weighted additive reward function and then using single-objective value-based methods such as Q-learning or policy search-based methods such as A2C for handling multiple objectives \cite{ruadulescu2020multi, roijers2013survey}. Naturally, this approach leads to a single-policy solution since the underlying single-objective methods are designed to find a single optimal solution. 
The techniques that have used a weighted reward function, although capable of incorporating multiple objectives, do not have the same flexibility as an explicitly multi-objective system that includes multiple agents \cite{hayes2022practical}. Determining the right weights to assign for each of the objectives requires domain expertise, and even the guesses made by experts are not guaranteed to be optimal. Furthermore, with a weighted reward function, a change in preferences cannot be accommodated flexibly.  It is also costly to run multiple combinations for identifying the most appropriate distribution of weights. 

\subsubsection{Multi-Objective RL (MORL)}
Multi-objective RL though inherently more complex to design as well as train compared to their single objective counterparts \cite{canese2021multi}, may be essential for efficiently capturing the diverse conflicting requirements that should be satisfied by workflow schedulers when formulating scheduling decisions in the cloud and edge computing environments. In \cite{hayes2022practical}, authors argue that designing an explicitly multi-objective system from the beginning leads to the reduction of computation time and sample complexity. Particularly in the presence of highly dynamic conditions such as those in the cloud and edge computing environments, it is impractical to assume that the desired trade-offs between objectives will remain constant. RL frameworks designed explicitly in a multi-objective manner are better suited for such environments since the agents can be trained so that they can adapt to such changes, this eliminates the need to retrain the model to identify a different policy every time an external factor causes a change in objective preferences. 

\subsection{Taxonomy based on agent architecture}

In this section, we analyse the Agent Architecture branch of the taxonomy shown in Figure \ref{fig:taxonomy}. When designing an RL-oriented solution, complex scheduling problems may require coordination among multiple agents (cooperating or competing) whereas a single-agent architecture may be more suitable for other more general scenarios. Despite the advantages of multi-agent methods, they do have certain limitations such as hindrance to learning which may occur due to the actions of multiple concurrent agents making the environment non-stationary and curse of dimensionality due to exponential growth in joint state and action spaces with the number of agents \cite{canese2021multi}.

\subsubsection{Single-Agent Systems} As the name implies, a single agent system comprises of a single RL agent that interacts with the environment and performs an action for which a reward is received. Single-agent systems are less complex to design as well as train since they are free from the common challenges associated with multi-agent coordination described above. Despite the success of single-agent RL methods, their applicability to a large number of real-world problems is limited since they cannot be solved with a single agent \cite{canese2021multi}.

\subsubsection{Multi-Agent Systems (MAS)} 

In multi-agent systems, multiple agents interact with the environment and learn to solve a problem concurrently. Multi-agent RL can be categorized into three groups based on whether the nature of interactions between agents is cooperative or competitive. \\

\textbf{Cooperative multi-agent systems} RL based schedulers of this category leverage the cooperation among multiple agents for solving parts of a larger problem. 
The agents may work independently on different parts of the problem or in parallel on parts of a single problem distributed among them \cite{mammen1998problem}. The concurrent processing of sub-problems by multiple agents enhances the scalability of the system which is particularly important for larger problems. Also, the use of multiple agents working in parallel makes the system more robust since the system is tolerant to the failure of a single agent.
For instance, in \cite{asghari2020cloud} both aforementioned types of multi-agent co-ordinations are leveraged to design an integrated task scheduling and resource provisioning with RL. In the task ordering phase, multiple parallel agents are used for efficiently analyzing parts of a workflow, and ordering the tasks such that tasks with higher ranks can be prioritized in the resource provisioning phase. Since different agents analyze different parts of a workflow, the search space handled by each agent is reduced, leading to faster learning. In cases where the same part of the workflow is analyzed by multiple agents, 
the weighted average of the Q value estimates of agents is taken. The use of parallel agents in this manner has been particularly useful for speeding up the learning process especially when workflows with a large number of complex interdependencies are involved. In the next phase of the same work, multi-agent coordination is again leveraged for provisioning resources to the ordered tasks. In this case, an agent is assigned to each resource, and the action is the selection of a task to be executed on the resource. Each agent operates with the objective of maximizing its reward, so to ensure the global objective of system-wide cost and energy minimization is achieved, a Markov game is formulated to converge the agents to a globally optimal solution.

\textbf{Competitive multi-agent systems} Competitive multi-agent systems are generally modelled as zero sum Markov games where the sum of returns of all agents is zero \cite{zhang2021multi}. In the common case where there are two competing agents \cite{littman1994markov}, the reward received by one agent is equivalent to the loss of the other. 

\textbf{Mixed cooperative-competitive multi-agent systems} Mixed settings are characterized by a combination of features from competitive and cooperative settings and includes a general-sum reward \cite{canese2021multi}. In mixed settings agents are self-motivated and may have rewards that are conflicting with those of others \cite{zhang2021multi}.

\subsection{Taxonomy based on the RL agent training and execution architecture}

In this section, we analyse the Training and Execution Architecture branch of the taxonomy shown in Figure \ref{fig:taxonomy}.
Multi-agent coordination in RL is an active research area, therefore the categorizations considered in this part of the taxonomy are not exhaustive. We have limited the scope of analysis to three main types of training and execution methods although other hybrid variants are recently being explored for multi-agent coordination \cite{santos2022centralized}.

\subsubsection{Centralized Training and Execution}

A majority of RL-based algorithms considered in this survey belong to the category of centralized training and centralized execution as they are straightforward single-agent algorithms. As the name implies, in the case of a single-agent scenario, the agent is trained and executed in a centralized manner. In multi-agent scenarios, this type of learning is typically based on a joint action and observation model in which a joint observation of all agents is mapped to a joint action by a centralized policy. This in turn could lead to an exponential growth in the action and observation spaces with the number of agents \cite{gupta2017cooperative}. The multi-agent workflow scheduling algorithm proposed in \cite{jayanetti2022deep} leverages hierarchical RL  \cite{barto2003recent} for addressing the aforementioned problem. In the proposed approach, the action space is designed in a hierarchical manner such that the action spaces of agents at each level are confined.

\subsubsection{Centralized Training and Distributed Execution}

Despite the promising experimental results, centralized execution-based RL algorithms may not be ideal for highly distributed and stochastic edge computing environments \cite{tuli2020dynamic}. In centralized training and distributed execution paradigm, additional information 
that is only available at training time is leveraged to learn decentralized policies such that during execution time the agents operate solely based on local observations and partial information about the intentions of other agents, thus alleviating the need for complex communications between agents \cite{foerster2016learning}. The paradigm of CTDE is utilized in \cite{zhang2023multi} for coordinating multi-agent training and execution in the scheduling of concurrent requests modeled as DAGs across edge networks. A policy learning technique based on value decomposition networks \cite{sunehag2017value} is adopted for decomposing the value function of the team into individual value functions. This helps overcome problems associated with independent learning such as when the learning of a particular agent is discouraged since its exploration leads to the hindrance of another agent that has already learned a useful policy \cite{sunehag2017value}.

\subsubsection{Distributed Training and Execution}

As the name implies, in the paradigm of distributed training and execution, both learning as well as execution takes place in a decentralized manner. Popular Multi-Agent RL algorithms such as A3C \cite{mnih2016asynchronous} and more recent developments such as IMPALA \cite{espeholt2018impala} as well as algorithms such as Gorila (General Reinforcement Learning Architecture) \cite{nair2015massively} that leveraged parallel computations to enhance the efficiency of single-agent training falls into this category. 

\subsection{Taxonomy based on scheduling objective}\label{section-objective}

In this section, we analyse the Scheduling Objective branch of the taxonomy shown in Figure \ref{fig:taxonomy}.

\subsubsection{Makespan}

Makespan minimization is one of the most commonly studied objectives in workflow scheduling across distributed computing infrastructures. Makespan refers to the total time it takes for a workflow to complete execution, and it is dependent on multiple factors including the size of tasks, processing speeds of nodes to which tasks are allocated for execution, the volume of data dependencies among tasks, and bandwidth of underlying networking infrastructures. Makespan minimization across heterogeneous and distributed infrastructures is a complex problem, it is even more complicated when workflows are to be scheduled across multi-tenant cloud computing environments \cite{hilman2020multiple}. Several RL-oriented methods have been proposed by researchers for solving this problem efficiently. One such approach is to use RL for sorting tasks of workflows prior to resource provisioning (see section 3.1.4), another more common method is to incorporate makespan minimization objective to the reward of an RL agent so that resources are allocated to tasks with the long term objective of minimizing the overall makespan.  

\subsubsection{Energy efficiency}

The ever-increasing popularity of the cloud computing paradigm has resulted in the development of hyper-scale data centers that consume significantly high levels of energy. High energy consumption leads to undesirable consequences including raised levels of CO2 emissions and increased monetary costs associated with high electricity consumption \cite{avgerinou2017trends}. Energy efficiency is equally important for the emerging edge computing paradigm since a majority of edge devices are likely to be powered with batteries with limited capacities and the collective energy consumption of rapidly growing volumes of edge nodes is estimated to be significantly high \cite{jiang2020energy}.  
Therefore, a plethora of research efforts has been focused on enhancing the energy efficiency of cloud and edge computing infrastructures through a variety of different mechanisms including dynamic power management \cite{benini2000survey}, thermal aware scheduling \cite{ilager2020thermal} and renewable energy utilization \cite{deng2014harnessing}. Accordingly, energy efficiency has been considered a primary objective in a large number of workflow scheduling algorithms proposed in academia \cite{medara2022review}. More recently, researchers have leveraged the advanced capabilities of RL for enhancing the energy efficiency of workflow executions across distributed cloud computing environments.

\subsubsection{Cost}

A majority of cloud services are billed using the pay-as-you-go approach. Cloud computing platforms offer virtual machines (VM) with different flavors (processing capabilities including the number of virtual CPUs, memory, and storage) and therefore different prices are charged for their utilization. The total execution cost of a workflow depends on the total number of VMs used for the execution of workflow tasks and their respective flavors, which is determined by the underlying resource provisioning and scheduling policy. In addition to minimizing overall cost as a primary objective, some studies have attempted to satisfy budget constraints in a best-effort manner, while primarily optimizing different objectives such as makespan and/or energy \cite{qin2020energy}.

\subsubsection{Resource Utilization} Optimized resource utilization is an intrinsic objective that goes in hand with many other objectives. For instance, optimized network and host (virtual or physical) resources lead to lower energy-efficiency, since unused resources can be put into dormant state. This in turn reduces cost of operations as well. 

\subsubsection{SLA/Deadline} Deadline is an important objective particularly when it comes to delay sensitive workflows such as those used in IoT (Internet of Things). In some studies deadlines are also used for establishing an upper bound to the degree to which makespan of workflows are allowed to extend for achieving objectives such as energy-efficiency \cite{jayanetti2022deep}. 

\afterpage{
\begin{landscape}
\begin{table}[!htbp]
	\caption{Analysis of existing literature based on the taxonomy for application placement policy [CTE - Centralized Training and Execution, CTDE - Centralized Training and Distributed Execution, DTE - Distributed Training and Execution, MORL - Multi-Objective Reinforcement Learning, SORL - Single-Objective Reinforcement Learning]}
	\label{table:tax2Works}
	\resizebox{\linewidth}{!}
	{\renewcommand{\arraystretch}{1.7}
	\begin{tabular}{l c c c c c c}
			\toprule
			\multicolumn{1}{c}{\textbf{Work}} 
   &\multicolumn{1}{c}{\textbf{Agent Action}} 
   &\multicolumn{1}{c}{\textbf{RL Algorithm}}
        &\multicolumn{1}{c}{\textbf{RL Objective}}
			&\multicolumn{1}{c}{\textbf{Agent Architecture}}
			&\multicolumn{1}{c}{\textbf{Training-Execution Architecture}}
			&\multicolumn{1}{c}{\textbf{Scheduling Objectives}}
			\\
			\hline
			A. Asghari et al \cite{asghari2020cloud} & Task Sorting & Q Learning & MORL & Multi-Agent Cooperative & CTDE & Makespan, Energy-Efficiency \\
            & Task Admission & & & & &  Cost, Resource Utilization \\
			\hline
   Y. Wang et al \cite{wang2019multi} &  Execution Node Selection & Deep Q Learning & MORL & Multi-Agent Cooperative & DTE & Makespan, Cost   \\
   \hline
   A. Asghari et al \cite{asghari2021task} & Task Sorting & SARSA & MORL & Multi-Agent Cooperative & CTDE & Resource Utilization, Makespan \\
   & Task Admission & & & & & \\
   \hline
   Y. Qin \cite{qin2020energy} & Execution Node Selection & Q Learning & MORL & Single-Agent & CTE & Makespan, Energy-Efficiency, \\
   & & & & & & Cost \\
   \hline 
   A. Nascimento et al \cite{nascimento2019reinforcement} & Execution Node Selection & Q Learning & SORL & Single-Agent & CTE & Makespan \\
   \hline
   A. Asghari et al \cite{asghari2020online} & Task Sorting & Q Learning & MORL & Multi-Agent Cooperative & CTDE & Makespan, Energy-Efficiency, \\
   & Task Admission & & & & & Resource Utilization \\
   \hline
   Z. Tong et al \cite{tong2020ql} & Task Sorting & Q Learning & SORL & Single-Agent & CTE & Makespan \\
   \hline
   A. Kintsakis et al \cite{kintsakis2019reinforcement} & Execution Node Selection & REINFORCE & SORL & Single-Agent & CTE & Makespan \\
   \hline
   T.Dong et al \cite{dong2020task} & Execution Node Selection & Deep Q Learning & SORL & Single-Agent & CTE & Makespan \\
   \hline
   Z. Pheng et al \cite{peng2020multi} & Execution Node Selection & Deep Q Learning & MORL & Single-Agent & CTE & Energy-Efficiency, Makespan \\
   \hline
   A. Orehan et al \cite{orhean2018new} & Execution Node Selection & Q Learning, SARSA & SORL & Single-Agent & CTE & Makespan \\
   \hline
   Q. Wu et al \cite{wu2018adaptive} & Execution Node Selection & REINFORCE & SORL & Single-Agent & CTE & Makespan \\
   \hline

   Y. Hu et al. \cite{hu2019learning} & Task Admission & Actor-Critic & SORL & Single-Agent & CTE & Makespan \\
   \hline
H. Li et al. \cite{li2022weighted} & Composite Action & Double Deep Q Learning & MORL & Multi-Agent Cooperative & CTE & Makespan, Cost \\
 & (Task Admission, Execution Node Selection) & & & & & \\
\hline
F. Hue et al \cite{xue2022deep} & Execution Node Selection & Deep Q Learning & SORL & Single-Agent & CTE & Makespan \\
\hline
Y. Zhang et al. \cite{zhang2023multi} & Task Admission & Deep Q Learning &  MORL & Multi-Agent Cooperative & CTDE & Energy, Delay, \\
& & & & & & Throughput \\
\hline
Y. Zhang et al. \cite{zhang2020online} & Execution Node Selection & TD-Learning & MORL & Single-Agent &  CTE & Energy-Efficiency, Delay \\
\hline
Z. Hu \cite{hu2019spear} & Task Sorting & Deep Neural Networks & SORL & Single-Agent & CTE & Makespan \\
\hline
Z. Tong \cite{tong2020scheduling} & Execution Node Selection & Deep Q Learning & SORL & Single-Agent & CTE & Makespan \\
\hline
A. Jayanetti et al. \cite{jayanetti2022deep} & Execution Env Selection & Proximal Policy Optimization & MORL & Multi-Agent Cooperative & CTDE & Makespan, Energy-Efficiency \\
& Execution Node Selection & & & & & \\
			\bottomrule
	\end{tabular}
	\renewcommand{\arraystretch}{1}}
\end{table}
\end{landscape}
}

\section{Review of Reinforcement Learning based workflow scheduling techniques}\label{survey-review}

In this section, we review the RL-based workflow scheduling techniques in the context of the taxonomy presented in section 3. 

In \cite{wang2019multi} the problem of scheduling workflows in cloud computing is handled with the optimization objectives of minimizing makespan and cost. For this, the authors have proposed a multi-agent deep reinforcement learning framework. The multi-agent collaboration is modeled as a Markov game with a correlated equilibrium. As opposed to Nash equilibrium, in a correlated equilibrium, the agents are not motivated to deviate from the 'joint distribution in a unilateral manner' hence they are capable of collaborating together to optimize different objectives. Each agent attempts to optimize one of the scheduling objectives. It is assumed that the actions and rewards of agents are visible to other agents. The state of the system includes VMs that are available for execution currently, and the successors of tasks that have been scheduled for execution in the previous state. The action space consists of the probabilities of a task to be allocated to a particular VM. The reward of the makespan agent and the cost agent is designed to encourage the minimization of makespan and cost, respectively. 
 		
A DRL-based task scheduling and resource provisioning framework for workflow execution in the cloud is proposed in \cite{asghari2021task}. In the task scheduling step, an agent is assigned to each workflow for evaluating the values of nodes after a number of episodes. In this case, each node of the workflow is a state and an action is the selection of the next node (successor). A reward that depends on the computation cost of the next node and the communication cost between the current node and the next node is received by the agent and the values in the Q table are updated accordingly. The agent uses SARSA algorithm and after a number of iterations, the maximum cost from the start node to all other nodes is obtained. The tasks are then sorted in ascending order of their start times. In the resource provisioning phase, an agent is assigned to each resource. A state in this case is the sequence of tasks that was the result of the previous phase. An action corresponds to the selection of a task to be executed on the relevant resource. The reward received by the agent is the resource utilization of the resource post to the allocation of the selected task to the resource. Since the agents are operating independently, this method formulates the reward in a manner such that long-term resource utilization is optimized rather than sub-optimal greedy allocations. To ensure model convergence, and the selection of a globally optimal solution GA is used. Accordingly, the assignment of tasks to a resource is represented as a chromosome. Tradition GA operators (crossover, mutation, etc.) are then used for achieving model convergence. The results of a comparison study on a simple use case clearly demonstrate the superiority of using cooperative agents over the random selection and independent agents.

An end-to-end RL-oriented framework for resource provisioning and scheduling workflows in cloud computing environments is proposed in \cite{asghari2020cloud}. The first RL model is for ordering the execution order of tasks in a workflow such that the scheduling algorithm will pick tasks for scheduling from the ordered list. For this, a multi-agent RL framework is used, where multiple agents are assigned to a workflow, thereby reducing the search space of the problem for faster convergence. A state of the environment is a node of the workflow, and the state space comprises the set of all nodes of a workflow. An action includes selecting a child node and the reward is the sum of the current execution time of the node and the data transfer time to the selected child node. Where multiple agents have traversed through the same path, the value of a node is the average value of all the agents. In the resource provisioning phase, tasks are assigned to a cluster of resources. Again RL is used to determine which tasks are assigned to which resources in the cluster. In this model, a markov game is formulated where each agent is a resource and each action corresponds to selecting a task to be executed on the resource. A local reward based on the resulting allocation's ability to meet task SLA and the state of the resource (normal, overloaded, under-loaded) post to the assignment is given to each agent. Virtual machine migration and DVFS techniques are is incorporated into the resource provisioning scheme to further enhance the energy efficiency of the system. 

In \cite{qin2020energy} authors aim to minimize the makespan and energy consumption of workflows within a budget constraint. Tasks in a workflow are sorted based on a priority value calculated considering the task execution time and communication dependencies. The sorted tasks are then scheduled using Q learning algorithm. The agent environment considers VM utilization at each time step as the current state and an action corresponds to the selection of a VM for task execution. A budget constraint is imposed on the action space to limit available actions at each time step. A multi-vector reward in which one vector is the ratio of fastest and actual finish times of  task and the other vector is the ratio of least and actual energy consumption of task execution is received by the agent. Since the reward consists of two vectors (a weight selection problem arises), this work uses the Chebyshev scalarization function to secularise the Q values of state-action pairs and then selects the smallest scalarized Q value in a greedy manner. At the end of each episode, the corresponding solution is added to the Pareto set if it isn't dominated by any other solutions, and all solutions that are dominated by it are removed. 

In \cite{nascimento2019reinforcement}, a unique approach is proposed where Q learning is used to schedule tasks which are referred to as activations, to VMs with the objective of minimizing the makespan. However, different from existing works, in this approach, an episode corresponds to scheduling the activations of a single workflow. Hence the states are limited to available, unavailable, successfully finished, and terminated with failure. The action space only comprises of two actions which include scheduling an activation to a VM or doing nothing. A reward based on the performance of an assignment of activation to a VM compared to the overall performance of the workflow is used for promoting actions that improve the efficiency of the workflow. 

The proposed DRL framework in \cite{asghari2020online} is somewhat similar to that of\cite{asghari2021task}. The main difference is that work used SARSA whereas this used Q learning. Initially, DRL agents are assigned 'parts' of workflows, and they traverse the workflow to find the cost from each node to the sink node. The agents are rewarded based on the sum of the computation cost of the next node and the communication cost for sending data between the nodes. The sorted tasks are then allocated to resources using multiple RL agents each of which is assigned to a resource, and the agents operate with the objective of improving long-term resource utilization. If the task deadline is exceeded in a given resource, then the corresponding agent receives a large penalty thus encouraging agents to select tasks that leads to better resource utilization while also meeting deadlines. The reward for achieving the aforementioned goals consists of two components; the frequency of the processor to which the task is assigned and the negative value of the remaining unused frequency of the resource post to task allocation. The two components are combined through a normalized weight factor which can be used for adjusting the priority according to system requirements.  As an added advantage of the increased resource utilization, higher energy savings are also achieved. 

In \cite{tong2020ql}, a static task scheduling algorithm is proposed. It uses a combination of Q learning together with the popular HEFT algorithm for obtaining optimal task ordering. For each workflow, an entry task is randomly chosen which is considered the current task (current state), one of the previously unselected successor tasks is then selected and an immediate reward is calculated which is equivalent to the upward rank proposed in the HEFT algorithm. Q table is updated accordingly. The selected successor then becomes the next state. The process repeats until the Q table converges. The final Q table is then used to obtain the optimal task order. For the execution of a task, the processor which is capable of completing the task earliest is selected. 

In \cite{kintsakis2019reinforcement}, supervised learning techniques are used for predicting the probability of failure and runtime estimations of tasks at different execution sites. These predictions coupled with the cost of communicating input data to a particular site and the number of task successors are formulated as a feature vector. For each of the ready tasks, and for each of the available execution sites, such vectors are constructed and all of these together form the input sequence. An action would be the assignment of ready tasks to execution sites, and the work assumes that a task can be assigned to one site and a site can only execute one task at a time. The size of the action space would then be equivalent to the number of execution sites that are capable of executing a task. For handling the input and output sequences which are of variable size a pointer network is used in this work. The size of action space is reduced by formulating the output of the pointer network to select one task to execution site allocation at a time. The model is executed until all ready tasks are assigned to sites for execution. The reward is equivalent to the negative workflow makespan and from this, a baseline of the average execution time of a workflow based on past observation is deducted to stabilize the training process. The proposed model is trained in a simulated environment and deployed and tested in a practical environment. 

\cite{dong2020task} presents a straightforward application of DQN for scheduling tasks with precedence relations in a cloud manufacturing environment. Tasks are sorted prior to the use of DQN for scheduling using upward ranks. State space comprises the server workloads and tasks allocated to servers and the tasks' start and finish times. The action space consists of all the servers to which a task can be allocated and the makespan difference between the current and next state forms the reward. 

The work proposed in \cite{peng2020multi} uses DQN with a weighted reward function for establishing a desired tradeoff between energy consumption and makespan in scheduling tasks with precedence relations. More specifically, the makespan component of the reward is the inverse of the total wait and execution times of a node at the selected server and the energy component is the difference in energy consumption between the current and previous time steps. Min-max normalization was used to normalize the two components prior to their application of them in the weighted reward function. The state space comprises the number of VMs available in servers and the waiting time at each server for a task to be deployed, the set of servers available for task execution forms the action space.

In \cite{orhean2018new}, authors proposed an RL framework for scheduling workflows in distributed computing environments. A multi-threaded java based pluggable scheduling module is presented such that multiple clients can be served by leveraging the parallel processing capabilities. The authors have implemented Q-learning and SARSA algorithms in the presented module. The scheduling environment is designed such that the state space consists of the load level of server queues defined in relation to a precision percentage, and an indication of whether a predecessor or a sibling of the task already resides in a particular server queue. An episodic reward of total execution time in comparison to a base value is awarded to the agent at the terminal state. As opposed to most workflow scheduling papers which have simply failed to include an indication of the node in which a particular task's predecessors and successors are residing, this work has incorporated that information into the state space, and that in turn enables the agent to learn to make better allocations such that the resulting allocations result in lower makespans due to minimized communication times and improved parallel executions. Since this work has used, Q-learning and SARSA it is important to prevent the expansion of state space, and that is achieved through the use of a precision percentage to indicate the number of tasks in a queue and also, rather than including the specific characteristics of the task to be scheduled, a task classifier is used to assign a task type to each task. This type of state space discretization although lowers accuracy is crucial when algorithms such as Q-learning and SARSA are used. 

\cite{wu2018adaptive} uses REINFORCE algorithm to schedule precedence-constrained tasks in distributed computing environments. To formulate the state space in a compact manner, they have incoporated the earliest start time of a task in each of the available servers. The earliest start time serves as an indication of both the load on the processor as well as the cost of communication. Additionally, the number of tasks that are to be scheduled is also included in the state. Where multiple tasks are ready to be scheduled, upward ranks are used for prioritizing the selection of tasks. An immediate reward of the increase in schedule length after taking the current action compared to the previous schedule length is used.

The authors of \cite{hu2019learning} use images for state-space representation, similar to the popular job scheduling framework presented in \cite{mao2016resource}. The resource availability and usage of the cluster together with resource requirements of the tasks to be scheduled and the number of scheduled tasks are included in the state. In order to include inter-task dependencies in the state, such that the agent is capable of learning better, a critical path-based technique is proposed. In the proposed approach, a workflow is divided into multiple stages (the number of stages equals to the number of tasks in the critical path) and processed with a depth-first search. Using the critical path information computed during the division process, a stage number matrix and a critical path matrix are computed for each task and these matrices are also included in state representation. An action corresponds to the selection of a task to be scheduled in the cluster and in one time step multiple such selections are permitted to prevent the action space from being too large. The agent is also allowed to select a movement action that causes the system to run one more time-step (without admitting a new task to the system) An episodic reward equivalent to N/makespan where N is the number of tasks scheduled is awarded to the agent. 

In \cite{li2022weighted}, authors propose a multi-level multi-agent reinforcement learning framework for scheduling workflows in cloud computing environments with the objectives of minimizing makespan and cost. Furthermore, the paper also presents a mechanism for adjusting the models' attention to each objective as preferred by users during the training process so that the diversity of the resulting solutions is enhanced. The state space includes the features of ready tasks, VMs available as well as corresponding time and cost of executions. In the multi-level scheduling strategy, first, a task is selected from amongst the ready tasks and then a VM is selected for executing the selected task. Since the number of ready tasks is variable, a pointer network is used in the first level so that an input state with variable length can be handled. The second level uses a traditional deep neural network. For each of the objectives, a separate sub-agent each with a separate reward is used at each level. The work also proposes the use of a normalized weight factor for combining the probabilities of selecting a candidate action for time and cost agents at each level (task selection and VM selection). The authors have also used a weighted double-deep Q learning network with a dynamic coefficient. Double deep Q learning (DDQN) method reduces overestimation of action values (associated with traditional deep Q learning method) and weighted double deep Q learning (WDDQN) further improves accuracy by reducing the underestimation bias associated with DDQN. To establish a desirable balance between over-estimation and underestimation issues associated with DQN and DDQN respectively, a dynamic coefficient that can be changed during the training process according to the two types of errors in estimations is proposed.

The authors of \cite{xue2022deep} combined DQN and genetic algorithm to design a scheduling algorithm with high convergence speed in an edge computing environment. More specifically, DQN is used to generate the initial population of the Genetic Algorithm and this in turn improves the convergence speed of the algorithm by eliminating the randomness of the initial population. The state space of the DQN model comprises the start and finish times of tasks in edge servers, cost of communication, and computations. An action corresponds to the selection of a server and the reward is the difference between makespan in the current and next states.  

In \cite{zhang2023multi}, a multi-agent DRL framework for scheduling DAG-based user requests in edge computing environments is proposed. In order to handle the problem of non-stationary environments that occurs when independent DQN agents are operating in an environment, a value decomposition network coupled with  Centralized Training and Distributed Execution (CTDE) training method is used. Accordingly, the authors use the linear summation of individual Q values of agents to derive the team Q value. As opposed to more complex ways of computing the team Q value with techniques such as neural network fitting, a linear summation simplifies implementation and also provides an intuitive evaluation of each agent's contribution to the global objective. Furthermore, to avoid the problem associated with non-stationary environments, the authors have trained the model in a Centralized Training and Distributed Execution manner (CTDE). Accordingly, all agent training tuples are stored in a shared replay buffer from which random samples are selected and used for training all agents through the replay. The state of each agent comprises of resource capacity of the node, communication cost to other nodes, and the details of ready tasks. Action is the joint actions of all edge nodes, and at each epoch, only the first N tasks are selected (one edge node can only select one task). The reward is the weighted sum of the average delay and energy consumption of edge nodes and a penalty that is dependent on the number of tasks that exceeded the temporal dependence.

\section{Future Research directions}\label{survey-future}

\subsection{Supporting multiple objectives with multi-policy RL algorithms} 

Due to the diverse requirements of stakeholders real world scenarios typically require the workflow schedulers to optimize more than one objective. The existing RL based scheduling algorithms convert multiple objectives into a single objective with a single scalar reward function. The function may include multiple weighted components each relating to a particular objective. There are multiple problems with this approach \cite{hayes2022practical}. Firstly, the decision of which weights are to be assigned for which objectives is a manual process that requires significant domain expertise. Even an educated guess made by a domain expert could deviate from the optimal solution that could've been achieved if all potentially optimal policies were evaluated to find the policy that results in the best trade-off between multiple objectives. Furthermore, a change in objective preferences of the underlying system would make the current policy obsolete hence requiring the single objective agent to be retrained. Finally, evaluating multiple reward functions to identify the right match tends to be a costly process in terms of computation time and sample complexity. Therefore, it would be beneficial to explore how multiple objectives can be explicitly incorporated into RL frameworks designed for scheduling workflows across distributed cloud computing environments. Multi-agent architectures in which each agent is solely responsible for optimizing a particular agent is one potential means of achieving this \cite{ruadulescu2020multi}.

\subsection{Designing Multi-agent RL solutions for complex scheduling problems}
A majority of existing workflow scheduling methods have proposed single-agent RL algorithms regardless of the dynamicity and decentralized nature of underlying infrastructures. Although single agent systems are less complex to design, they may not be efficient to capture the dynamics of more complex real world scheduling scenarios such as when workloads are to be scheduled across federated clouds and emerging fog computing environments. Accordingly, Multi-Agent Systems (MAS) in which interactions between multiple agents are leveraged, is proven to be a better fit for problem solving in highly distributed and stochastic environments compared to its single agent counterpart \cite{bucsoniu2010multi}. Therefore, it is worthwhile to investigate how multiple agents can be leveraged for designing more efficient RL solutions complex workflow execution environments. For instance, competitive multiple agents settings can be designed so multiple agents work concurrently on sub-problems of a larger problem which in turn enhances the scalability of the system as well as fault tolerance. 

Furthermore, as opposed to centralized cloud environments, in highly distributed multi-cloud and edge computing scenarios, it is likely that information about the status of all nodes are not centrally available in real time. It will be beneficial to use the centralized training and distributed execution paradigm in such scenarios so that the agents can be trained to learn decentralized policies with additional information that is only available at training
time. During execution the agents can operate solely based on local observations and partial information about the intentions of other agents. This enables more realistic modeling of the actual environment and also helps simplify
communications between agents \cite{foerster2016learning}.   

\subsection{Estimating task execution times accurately} 

Accurate estimations of task runtimes are important since some scheduling algorithms rely on such estimates for formulating scheduling plans. For RL based algorithms, such estimates are particularly useful since they can be used during the training process of agents to reward efficient allocations. The more accurate the estimations are, the better the learning of the agent will be. However, accurately estimating task runtimes is a non-trivial task particularly due to the performance variability that is inherent in cloud and edge computing environments. Factors such as interference due to co-located workloads from multiple tenants, geographical distribution of resources makes it difficult to produce accurate estimations. Few studies have used machine learning techniques for estimating run times \cite{pham2017predicting}. Techniques such as online incremental machine learning could be explored for achieving accurate task estimates.

\subsection{Using asynchronous RL methods for improving training efficiency }

The existing RL based workflow scheduling algorithms are mainly based on synchronous learning/training. However, in highly dynamic cloud environments, it is likely that the changing environments as well as stakeholder preferences give rise to the need for training and re-training RL models frequently. Therefore, it is beneficial to use asynchronous reinforcement learning techniques \cite{mnih2016asynchronous} for speeding up the training process without requiring the use of specialized hardware such as GPUs. In particular, the asynchronous advantage actor critic (A3C) in which multiple actor critic agents are asynchronously executed in parallel on separate instances of the environment, is highly effective in reducing training time. Importance Weighted Actor-Learner Architecture (IMPALA ) is a distributed agent architecture that is proven to achieve even better performance than A3C methods \cite{espeholt2018impala}. These methods can be leveraged by researchers for designing scheduling agents that are more data efficient and stable.  


\subsection{Handling large action spaces more efficiently}

Existing works have modeled the action spaces of  workflow scheduling problems with simple flat designs. This in turn results in large discrete action spaces, particularly if workflows are to be scheduled across large infrastructures such as multiple cloud datacenters. Such designs are more prone to Curse of dimensionality problem that inhibits the learning progress of RL agents. This issue is exacerbated in multi-agent settings due to exponential growth in joint state and action spaces with the number of agents \cite{canese2021multi}. Therefore it is worthwhile to explore techniques that result in more efficient action space designs. For example, in hierarchical reinforcement learning methods, the problem is sub-divided into a number of sub-problems \cite{barto2003recent}, and these methods can be used for designing action hierarchies. For instance, in multi cloud scenarios, action spaces could be subdivided at different levels including availability zone, subnet, rack and node levels. 
Action branching architecture is another potential direction that could be investigated in this regard \cite{tavakoli2018action}. Some studies have also used clustering techniques such as nearest neighbors \cite{de2011robust}, \cite{dulac2015deep}, and it is worth exploring how such approaches can be incorporated in workflow schedulers designed to operate across large-scale infrastructures.  


\section{Summary and conclusions}

Reinforcement Learning has emerged as a promising paradigm for dealing with highly dynamic and complex problems due to the ability of reinforcement learning agents to learn to operate in stochastic environments. More recently, well formulated deep reinforcement learning solutions have outperformed traditional methods in highly complex problems including robotics, and games such as Alpha-Go and Dota. Reinforcement learning agents are capable of interacting with the environments where they get exposed to the real-world dynamics and thereby build an internal model which is a more accurate representation of the operating environment. This in turn leads to self-adaptability and improved decision making amid changing conditions and uncertainties. Despite the benefits, there are multiple challenges associated with the application of reinforcement learning techniques including multi-objectivity, curse of dimensionality, scalability and coordination. Off-the-shelf algorithms are unlikely to be efficient at overcoming the aforementioned challenges and therefore more problem specific novel reinforcement learning techniques need to be formulated. 

In this paper, we reviewed the state-of-the art of workflow scheduling algorithms with reinforcement learning in cloud and edge computing environments. Based on the analysis we identified the merits and weaknesses of existing works, and potential areas of improvements along with some of the latest developments in the field of reinforcement learning that can be pursued by the research community.

\bibliographystyle{ACM-Reference-Format}
\bibliography{refs}
\end{document}